# Computing With Contextual Numbers

**Vahid Moosavi**                                                                                                                      SVM@ARCH.ETHZ.CH
*Chair for Computer Aided Architectural Design
(CAAD) and Future Cities Laboratory, ETH Zurich,
8092 Zurich, Switzerland*

**Abstract**

Self Organizing Map (SOM) has been applied into several classical modeling tasks including clustering, classification, function approximation and visualization of high dimensional spaces. The final products of a trained SOM are a set of ordered (low dimensional) indices and their associated high dimensional weight vectors. While in the above-mentioned applications, the final high dimensional weight vectors play the primary role in the computational steps, from a certain perspective, one can interpret SOM as a nonparametric encoder, in which the final low dimensional indices of the trained SOM are pointer to the high dimensional space. We showed how using a one-dimensional SOM, which is not common in usual applications of SOM, one can develop a nonparametric mapping from a high dimensional space to a continuous one-dimensional numerical field. These numerical values, called *contextual numbers,* are ordered in a way that in a given context, similar numbers refer to similar high dimensional states.

Further, as these numbers can be treated similarly to usual continuous numbers, they can be replaced with their corresponding high dimensional states within any data driven modeling problem. As a potential application, we showed how using contextual numbers could be used for the problem of high dimensional spatiotemporal dynamics.

Keywords: self organizing maps, representation learning, dimensionality reduction, encoding

## 1. Introduction

Self Organizing Map (SOM) is a generic methodology, which has been applied in many classical modeling tasks such as visualization of a high dimensional space (Vesanto, 1999), clustering and classification (Ultsch, 1993; Vesanto & Alhoniemi, 2000), prediction and function approximation (Barreto & Araujo, 2004; Barreto & Souza, 2006) across many diverse domains of application (Kohonen, 2013). Also, there are different extensions to the original algorithm that was introduced by Kohonen (1982). As a result of these diverse sets of applications and different versions of the original algorithm, depending on the perspective, there are different interpretations for SOM algorithm. For example one can compare SOM with other clustering methods or to compare it with space transformation and feature extraction methods such as Principal Component Analysis (PCA) (Yin, 2008). Also, It is possible to explain and compare SOM with vector quantization methods (Bishop et al. 1998; Kohonen 1990). Further, it is possible to explain SOM as a nonlinear function approximation method and to compare it with other neural network methods and radial basis functions (Barreto & Araujo, 2004; Barreto & Souza, 2006). However, in this work, based on a different way of looking at SOM, which is close to its explanation as a dimensionality reduction method (Luttrell, 1990, 1994), we propose a new functionality of SOM (as an encoder), in which any high dimensional input vector is mapped to a numerical value.

From a geometric or a topological point of view, it is common to consider SOM among nonlinear dimensionality reduction methods by which a high dimensional space will be transformed to a lower dimensional space (usually to a two dimensional grid) (Kohonen 1995). In general, based on the mechanism used for dimensionality reduction, methods differ from each other. For example, in PCA one finds a linear operator (in terms of linear algebra), which is defined globally, considering the variations across different dimensions of the original observed data.

From this point of view, original dimensions (features) of the data are the primary elements of transformation, constructing a representational space and the latent dimensions are derived as a linear combination of original high dimensional space. Opposite to this global mapping, there is another approach for dimensionality reduction, known as topology preserving approaches, in which regardless of the global dimensionality of the observed data set, given a (dis-) similarity measure among individual data points, one can construct a self-referential coordinate system, by which each instance of the observation becomes a dimension for itself compared to all the other points. Rooted in Topological Data Analysis (TDA) (Zomorodian, 2007), methods such as Multidimensional Scaling (MDS) (Kruskal, 1964), Locally Linear Embedding (LLE) (Roweis & Saul, 2000), ISOMAP (Tenenbaum *et al. 2000*), and Mapper (Singh *et al.* 2007) are among this category of dimensionality reduction methods. From this dimensionality reduction point of view, SOM is among topology preserving methods, in which data from high dimensional space is mapped to a new low dimensional space, through a set of mechanisms to preserve the local topology as much as possible (Kiviluoto, 1996; Kohonen, 1982).

However, our main argument in this paper is that although the idea of dimensionality reduction seems very interesting, it usually ends to qualitative visualizations and exploratory analyses, trying to describe the underlying nature of the data set (Roweis & Saul, 2000; Tenenbaum *et al. 2000;* Singh *et al.* 2007). This is the common task in TDA methods such as Mapper (Singh *et al.* 2007), where the final step of the analysis is some sort of descriptions about the shape or the clusters of the data, represented in a graphical form (That is. by a simplicial complex).

The main goal of this work is to go beyond exploratory analysis and to develop a methodology toward operationalizing the new reduced dimensionality into the next steps of modeling and computation. From this point of view our work can be seen in the category of representation and feature learning as discussed in (Bengio *et al.* 2013), where the learned low dimensional features are inputs to another level of and integrated model.

In Section 2, we present SOM in terms of a mapping from a high dimensional space to a lower dimensional space, followed by some qualitative and fundamental comparisons with similar methods. Next in Section 3, we introduce the concept of *contextual numbers* and their potential functionalities, explaining how these contextual numbers can be constructed using a one dimensional SOM. In Section 4 we discuss general applications of contextual numbers and then in Section 5, we present the results of some experiments with real data sets. Finally we conclude our work referring to potential future directions.

## 2. SOM and Indexing a High Dimensional State Space

The SOM algorithm can be explained in terms of two simultaneous processes. We consider the training data set, $X = \{x_i, \ldots, x_N\}$ as a set of $N$ points in a $n$-dimensional space, $x_i \in R^n$, $i = \{1, \ldots, N\}$.

Assuming a low dimensional grid with $K$ nodes (or cells), we have a set of indices $y_j, j = \{1, \ldots, K\}$, each with an attached unique weight vector, $w_j \in R^n$. During the training phase, for each data point $x_i$ an index, $y_j$, will be assigned in a way that similar points in the high dimensional space (based on the selected similarity measure) will be given similar indices. In other words, If $||x_i - x_j|| < ||x_i - x_k||$, then $||y_i - y_j|| < ||y_j - y_k||$, where $||.||$ is the selected similarity measure.

Symmetrically, during the training phase, the weight vectors of the indices, $w_j s$, which have been given initial values will be adapted in a way to become similar to their assigned data points and further to be similar to other weight vectors in their neighborhoods in the grid. These requirements can be explained in terms of *competition* and *adaptation* mechanisms (Kohonen, 1995). Therefore, as the final output of the SOM, we expect to have a low dimensional index for each original data points and if two data points are similar in the high dimensional space, their corresponding indices should be in the same region of the two dimensional grid. Usually the competition and adaptation steps will be repeated enough to minimize the average dis-similarity between training data and their corresponding weight vectors, known as quantization error. In terms of representation learning methods quantization error is equal to reconstruction error in encoder/decoder systems.

From topology and neighborhood preservation point of view, conceptual idea behind the low dimensional representation of SOM is similar to methods such as LLE, MDS and other similar methods, but with a main difference that in SOM the mapping from original high dimensional space to lower dimensional space happens in a self organized and indirect way, based on the embedded mechanisms in the training algorithm, while in LLE for example, the final reduced dimensions are the results of a two step optimization process (Roweis & Saul, 2000). However, the convergence of SOM algorithm to an ordered set of indices has been proved for the case of one-dimensional SOM (Erwin et al. 1992; Cheng 1997; Flanagan, 2001).

On the other hand, as we will discuss in the following lines, SOM has several conceptual advantages over those topological methods, designed specifically for dimensionality reduction.

In majority of topological approaches to dimensionality reduction, there is always a one to one relation between original and latent representation of each data point (That is. $x_i$ and $y_i$ uniquely for data point $i$). Therefore, as it is pointed out in (Lum *et al.* 2013; Singh *et al.* 2007), there is no data reduction in traditional dimensionality reduction methods mentioned above. However, in SOM the number of indices (nodes) of SOM is not meant to be equal to the number of data points. When the number of indices is much smaller than the number of training data, which is the case in classical SOM, several data points will be assigned to the same index and therefore, each index acts like a data reducer and a cluster. In this sense, each index $y_i$ and its corresponding weight vector $w_i$ can be considered as the center of a possible state of the joint probability distribution over the original high dimensional data space and the fraction of training data points, assigned to that index, can be considered as the probability of occurrence for that composite high dimensional state (Kostiainen & Lampinen, 2002). From this point of view, one can argue that these indices are not necessarily low dimensional representations of the original data points, but instead considering the coexistence of the new low dimensional indices and their high dimensional weight vectors, SOM constructs a new symbolic space, which encapsulates the information of the original high dimensional space and these indices are pointers to the centers of those high dimensional probabilistic states.

In an opposite direction to data reduction, when the number of indices in the trained SOM is relatively large, which is the case of *emergent* SOM (Ultsch, 2005); there will be more than one index around a unique training data point. Thus, the final SOM creates a smooth pattern, filling the gaps between the training data points in the higher dimensional space. Consequently, SOM will be able to generalize from the training data points.

Further, it is worth to mention that unlike topological data analysis, where the final goal is to discover the nature (For example. shape) of the high dimensional data space (Lum *et al.* 2013), SOM indices and their weight vectors can be optimized in a fully targeted way. Since SOM is normally used for further steps such as function approximation or prediction of specific features, it can be implemented in a way that the final map and its indices is biased toward a certain goal, but not necessarily toward unfolding and discovering the natural topology of the original manifold. In other words, SOM can be considered as a generic instrument that produces a set of homogenous indices according to a defined target.

With this perspective to SOM, as a method for indexing high dimensional states, in the next section we will show how we can construct a set of indices, which can be used directly as numerical values.

## 3. SOM Indices as Contextual Numbers

SOM has been used in different classical modeling tasks such as data clustering (Vesanto & Alhoniemi, 2000), classification and prediction (Barreto & Araujo, 2004; Barreto & Souza, 2006). In all of these applications, a final index of the trained SOM will not be used directly as a numerical value, but through its assigned weight vector. For example, in the case of function approximation in combination of KNN method and SOM (Barreto & Araujo, 2004; Barreto & Souza, 2006), all the calculations and computations are performed based on the weight vectors of the SOM nodes, while the numerical values of the trained indices are of no numerical importance, but just as unique IDs or pointers. Therefore, comparing the approach of applying KNN algorithm on the original training data to the combination of KNN on SOM weight vectors, in the latter approach, SOM is used as a smoothing and de-noising filter over KNN. The same argument holds for the application of SOM for data clustering, when for example, one runs a clustering algorithm such as K-means over the weight vectors of the trained SOM instead of the original data (Vesanto & Alhoniemi, 2000). In none of these applications, SOM indices are used as numerical values.

The main idea in this work is that opposite to the current way of using SOM indices, if we design these indices in a way that they form a new one-dimensional space of indices, they can be used directly as some sort of numbers and numerical values. Therefore, for example one can assign an index with a numerical value to a high dimensional object such as an image or a segment of an audio or in principle any other representable object. From this point of view, SOM indices have similar function to low dimensional representation in autoencoder algorithm (Hinton & Salakhutdinov, 2006), but with a main difference that we aim to use final SOM indices as a continuous numerical field, while in methods such as semantic hashing (Salakhutdinov & Hinton 2009), the final low dimensional representation is a low dimensional binary code, being used in a look-up table.

The key point toward this goal is to use a SOM with a one-dimensional grid. In the following lines we argue how this is only possible in a one dimensional SOM.

However, the common choice of dimensionality in SOM grid is two in a form of a planar grid or as a toroid shape grid (Kohonen, 2013). The main reason for this choice is that a planar grid gives a very good way of visualizing a high dimensional space, by visualizing several dimensions (features of the original high dimensional space) with several colored maps at once, known as component plane (Vesanto, 1999) and psychologically it might be more understandable for non-expert people, as it can be read like geographical maps.

Figure 1 show the usual indexing of nodes in a two-dimensional and a one-dimensional SOM respectively. In both grids, the total number of indices is equal to $K$. In the planar grid, $C$ is the number of columns and $\frac{K}{C}$ is the number of rows.

As it is mentioned before, based on the adaptation mechanism in the training algorithm of SOM, indices in the same neighborhood of the grid have similar patterns in terms of their weight vectors. Therefore, assuming a Gaussian neighborhood around a specific index, we expect to have a distribution of similar nodes around a specific node as it shown in Figure 1 In a two dimensional grid the neighborhood similarity expands in two directions. Therefore, there is no direct correlation between the numerical values of indices and the similarity of their weight vectors. On the other hand, the simple, but interesting property of a one dimensional SOM is that there is a direct correlation between indices and the similarity of their weight vectors as follows:

If $\|y_i - y_j\| < \|y_i - y_k\|$ then $\|w_i - w_j\| < \|w_i - w_j\|$ (1)

Where, $1 \leq y_i \leq K$ is a scalar value, which is proved to be true for the one dimensional SOM (Erwin et al. 1992; Cheng 1997; Flanagan, 2001).

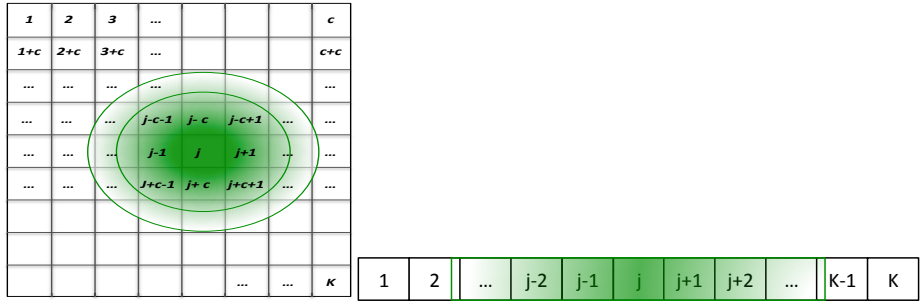

Figure1: Indexing of nodes in SOM with a two dimensional grid (left) and a one dimensional grid (right)

As an exemplary application, suppose that the training data points are one-dimensional random vectors with values in a specific range. Training a one dimensional SOM, we expect to see a direct correlations between indices and their weight vectors as both are one-dimensional vectors. As it is shown in 2 (left) the values of the weight vectors (here, scalar values) are correlated with their corresponding indices. On the other side, Figure 2 (right) shows the result of applying a two dimensional SOM. As we could expect, neighborhood similarity follows a bidirectional Gaussian distribution and according to the indexing method defined above, there is a cyclic correlation between indices and their corresponding weight vectors. Note that one might argue that with a diagonal indexing, starting from top left corner we would have a direct correlation with indices and their weight vectors, same as a one dimensional SOM. But the problem in that case is that the indexing is no more unique as there will be several nodes having the same distance to the top left corner or consequently this indexing should be reduced to a one dimensional SOM along with the diagonal of the two dimensional grid.

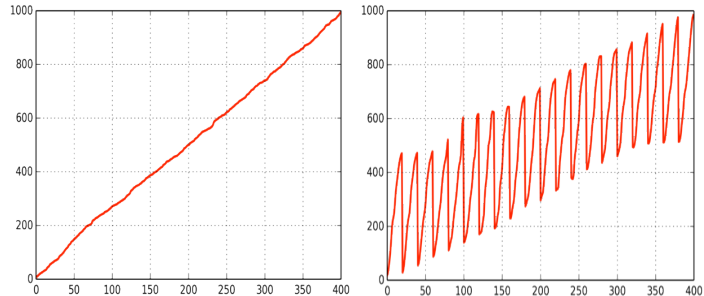

Figure 2: (Left) A one dimensional SOM with 400 indices trained by a set of 500 one dimensional data sets with random values ranging from 0 to 1000. (Right) A two dimensional SOM with 400 nodes in a grid of 20 by 20, applied to the same data set.

The interesting point here is that SOM itself does not know about the context in which the data have been observed, but it is able to find an ordered set of indices that encapsulates that context in a form of new numerical values. In general, if one changes the context of the data to higher dimensional spaces, this method should work.

In order to see whether the condition (1) holds for higher dimensional data sets, we trained a SOM by a set of spatiotemporal data from. Data set is a time series of Sea Surface Temperature (SST) for a specific region from (Rogers *et al.* 2013). Each data point in each time step is a matrix of numerical values with the dimension of 5 by 6. Figure 3, shows the pairwise Euclidean distance of the weight vectors of indices in trained SOM with 50 indices. As expected, neighboring indices are referring to similar high dimensional points (That is. have similar weight vectors and lower Euclidean distance).

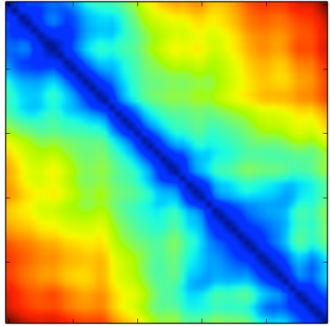

Figure 3: Pairwise Euclidean distance of weight vectors in one-dimensional SOM with 50 nodes, applied to a 30 dimensional data set.

Rendering each data point in this case as an image, Figure 4 shows how the final indices of the trained SOM encapsulate the prototypical SST patterns, smoothly changing from one side to the other side of the trained SOM.

Further, we applied a one dimensional SOM with 400 nodes to the face position data sets (Tenenbaum *et al.* 2000). Figure 4 shows two segments of the final SOM (indices from 1 to 20 and from 381 to 400), indexing high dimensional points (face position images) in an ordered manner.

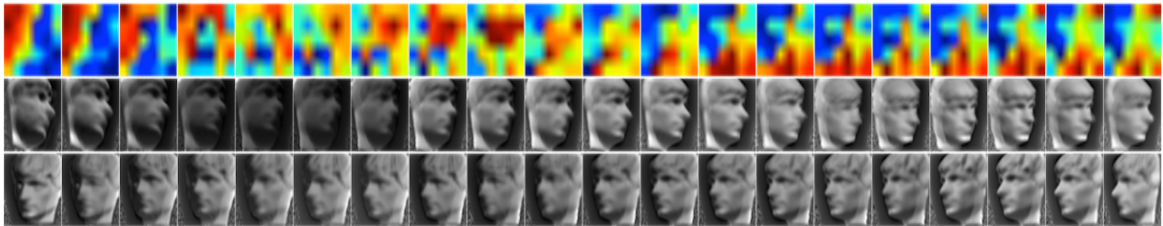

Figure 4: Rendering the weight vectors of a one dimensional SOM with 20 nodes trained by 5x6 dimensional data from (Rogers et al. 2013) (top row) and the final results of training a SOM with 400 nodes on the face position data set (Tenenbaum et al. 2000). Second row: indices from 1 to 20 and the third row showing the weight vectors of indices 381 to 400.

Therefore, a one dimensional SOM can be seen as a sequence of ordered numbers, pointing to a high dimensional space. However, comparing to classical numbers, such as natural numbers, the main difference is that here, the logic and meaning of ordered indices or signs (For example, 1, 2, 3…) are not presupposed, but found and symbolized from the contextual similarities within the training data. Thus, we call these ordered indices, ***contextual numbers***. One can claim that the contextual numbers are an abstraction from classical numbers expanding from one-dimensional order (which is the case in natural, rational or real numbers) to higher dimensional self-organized orders. In addition, the logic of the ordered set is not given a priori, but given implicitly from the context that training data have been observed.

The main interest to these contextual numbers is that they symbolize the original high dimensional data representations to scalar values, and while these scalars automatically keep the context, they can be used for further calculations. As we discussed in Section 2, from this point of view, SOM has a constructive approach comparing to traditional dimensionality reduction methods. The contextual number opens up higher levels of observation and calculations, while the ultimate goal of dimensionality reduction methods or those of topological data analysis is to investigate the nature of the observed data, ending to visual and qualitative findings.

### 3.1. Mapping and Reconstruction

Once a one dimensional SOM was trained, for mapping high dimensional data including both training and new data points, high dimensional points will be projected to the trained SOM. Using the softmax function, the posterior probability of assigning contextual number $j$, $cn_j$ to data point $x$ will be calculated as follows:

$$p(cn_j|x) = \frac{exp\left(s(w_j,x)\right)}{\sum_{j=1}^{k} exp\left(s(w_j,x)\right)} \quad (2)$$

Where $s(w_j,x)$ is the similarity function (or the neural activation), calculated as an inverse function of the distance between two high dimensional vectors of $x$ and the weight vector of index $j$, $w_j$.

Based on this posterior probability, there will be different choices to assign a contextual number to a given high dimensional data point. One approach is to use a weighted average of weight vectors over $g$ indices, selected indices with highest activations as calculated by equation 2, as follows:

$$cn(x) = \frac{\sum_{j=1}^{g} p(cn_j|x) cn_j}{\sum_{j=1}^{g} p(cn_j|x)} \quad (3)$$

Provided that the posterior probability distribution has a dominant peak, the final contextual number can be selected as follows:

$cn(x) = \text{Argmax}_j \left(p(cn_j|x)\right)$ (4)

Training a SOM, based on SST data used previously, Figure 5 shows the distribution of posterior probabilities of potential contextual numbers for two randomly selected data points, one from training data set (red) and one from new observations (blue). As we could expect, for both cases there is a distinctive maximum point, while the maximum probability for the new data is slightly lower than the maximum probability for the training data, which is expected as the SOM is optimized to represent the training data. However, having a very low peak value for the new data points could be an indicator to update and retrain the SOM with a new data set. We will discuss this issue in Section 5.

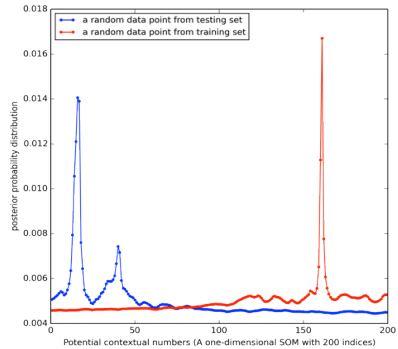

Figure 5: Distribution of posterior probabilities (neural activation), for assigning a contextual number from the continuous range of 200 indices for two random data points, one from training data (red) and one from a new observation data point (blue)

From state space modeling point of view, the contextual numbers group the high dimensional space into $k$ distinct clusters, with a specific property that similar clusters have similar indices (That is. similar numerical indices). Thus, the final SOM indices form a discrete field of integer values, each with a unique high dimensional weight vectors. In the case of weighted average over $g$ indices with the highest probabilities, if the final assigned contextual number is not an integer, one only needs to approximate its corresponding high dimensional vector by interpolating the weight vectors of its adjacent left and right integer contextual numbers.

In this work, we will use the indices as numerical values; however, from a representation learning point of view, the neural activation (or posterior probability distribution), which is shown in Figure 5 can be seen as a sparse representation (Bengio *et al.* 2013; Hinton & Salakhutdinov, 2006), in which only few features are activated for any input data. From this angle, SOM can be explained in terms of a hierarchical neural network with three layers, where the first and the final layers are the input and its reconstruction (That is. the weight vector of the SOM nodes), and in the hidden layer there are the indices of the trained SOM (Luttrell, 1990, 1994).

However, assuming SOM indices as a sparse representation, there is a main difference between this feature set and those of usual multilayer neural networks such as autoencoders.

The main difference is that in a multilayer neural network the feature set of each layer is a linear combination of its previous representational space such as $Y = WX$, while in SOM algorithm there is no direct transformation operator from a high dimensional features to all the SOM indices. Further, SOM learns its low dimensional representation (That is. its indices) in parallel with its unique high dimensional weight vectors.

The crucial point, which makes a main difference in two approaches is that how to interpret neural activation. In general, in deep networks, each index is a one dimension for itself and its neural activation shows the value of that dimension. Therefore, the whole index set can be considered as a $K$ dimensional feature (for a SOM with $K$ indices). On the other hand, in the case of contextual numbers, each index is a value in a one-dimensional representation and the neural activation is just indicating the probability of each potential value (That is. index) in this one-dimensional vector. However in the case of one dimensional SOM, since neighborhood indices refer to similar contexts (That is. similar weight vectors), the contextual numbers can be treated as a continuous numerical field, while in the first approach usually codes (as multidimensional vectors) are being used as pointers to specific categories, like the case of image retrieval (Hinton *et al.* 2006) or text classification (Hinton & Salakhutdinov, 2006).

Finally, we should note that although sparse coding is much more flexible for representation of high dimensional points, which is very helpful for those cases of categorical data, in this work we are interested in application of SOM indices as numerical values, where they can be used for those systems with no discrete categories, but with a continuous high dimensional state space, when one is interested to index a high dimensional space based on its own context.

In the next section, we will present one of the potential application areas of contextual numbers.

## 4. Applications of Contextual numbers

Since the contextual numbers can be treated similarly to other numerical values, they can be embedded in any data driven modeling tasks. One of the main generic applications could be the problem of function approximation, in which having a set of observable variables, one is interested to find a function (or a joint probability distribution) over a set of observed variables, $X: \{x_i, ..., x_n\}$ and the target feature $y \in R$ as follows: $y = f(X)$ or from a probabilistic point of view, the goal is to find the optimum estimator for $p(y|X)$.

Classically, $x_i$ and $y$ are numerical values, but now having contextual numbers one is able to perform system identification tasks where the observable elements are of higher dimensionalities. This is the case in many applications such as multimodal learning and representation (Weston *et al.* 2010), or multisensor data fusion and estimation problem (Waltz & Llinas, 1990), which have been discussed extensively in different fields of application.

As an example of multimodal representation, we refer to the problem of image annotation, where the goal is to find a joint probability distribution over a collection of images and their user specified tags (Weston *et al.* 2010). In this case, one idea could be to estimate the joint probability distribution on top of two sets of contextual numbers, one referring to the images and one to the texts. One obvious advantage of this approach would be that the problem of finding an appropriate image for a text or vice versa is transformed from a (multiclass) classification problem to approximation of a function in a continuous space, where texts and images are sorted out based on their own contexts and a function has been trained over the these two contextual numbers.

One specific application of contextual numbers that in this work we focused on is the problem of *spatiotemporal dynamics*, where one is dealing with a high dimensional dynamical space. In general in a dynamical system, assuming that the state of the system at time $t$ is dependent on $d$ previous states, we will have the following condition:

$p(x_t | x_{t-1}, \ldots, x_{t-d}) = p(x_t | x_{t-1}, \ldots, x_0)$   (5)

Where, if $x_t$ is a one-dimensional vector (For example. a specific stock price at time $t$, or the surface temperature in a specific spatial point at time $t$), the problem is of the category of univariate time series forecasting. However, in this work we are interested in those models, where in each time step, the observation, $x_t$, is a high dimensional observation vector. There are lots of problems that involve with spatiotemporal patterns.

From Linear Dynamical Systems (LDS) point of view (Roweis & Ghahramani, 1999), the well established solution to the problem of multidimensional time series modeling is the application of Kalman filter method, in which one transforms the multidimensional time series observations to a lower dimensional latent state space, while the latent states transit in a Markov chain. Using LDS approach, one needs to construct two linear operators, one for modeling the transition probabilities in the latent space and the other one for back and forth projections between observations (That is. the higher dimensional space) and the latent states (the lower dimensional space). Although, LDS approach works very well for low dimensional observations, as we observed in our experiments, it has computational difficulties in dealing with high dimensional observations, as it requires calculating iteratively the inverse of a large matrix (Horenko *et al.* 2008).

From a graph theoretical point of view, the problem of spatiotemporal dynamics is equal to the problem of predicting the behavior of a random field, with unknown dynamics, on a network of fixed structure (Shalizi & Shalizi, 2003).

Considering spatial and temporal relations, there will be two distinct approaches for dealing with a time varying random field. The first approach is to consider each point in space and time, $x_{t,i}$, separately and to find its best predictor among all the other previous points. However, not all the previous points are always necessary for prediction of a specific point in time and space. For example, the main underlying idea of past light cone proposed in (Goerg & Shalizi, 2012) is to find the optimal set of predictive points based on a logical assumption that if spatial influence propagates in a finite speed, new events have effects on the spatially neighborhood points, while for more spatially remote events to have effects, they must be older. However, it should be noted that if the time resolution of observations is not high enough, every point will affect all the next points and therefore the past light cone will be equal to all the previous points, where for high dimensional vectors it becomes computationally expensive as one needs to optimize thousands of parallel estimators (one for each spatial point) (Horenko *et al.* 2008).

The second approach would be to take the whole random field at time $t$ as one specific state, considering its high dimensional configuration, and to predict the whole high dimensional state at once.

Assuming that we are able to find corresponding contextual numbers for each high dimensional observation at each time step, Figure 6 shows, how a high dimensional time series model can be converted to a one dimensional time series, where in each time step, $cn_t \in R$ is a contextual number corresponding to high dimensional observation, $x_t \in R^n$. Considering all the potential configurations in a high dimensional vector, this approach seems to be impractical (Shalizi & Shalizi, 2003). However, as we discussed before, as a data driven method, the final contextual numbers are not necessarily all the potential combinations in the high dimensional space, but a set of ordered numbers, each of which pointing to the most likely high dimensional states. Further, as one can simply interpolate the values in between two integer contextual numbers, having enough number of trained indices, the final contextual numbers can be seen as a continuous one-dimensional numerical field referring to a unique high dimensional space.

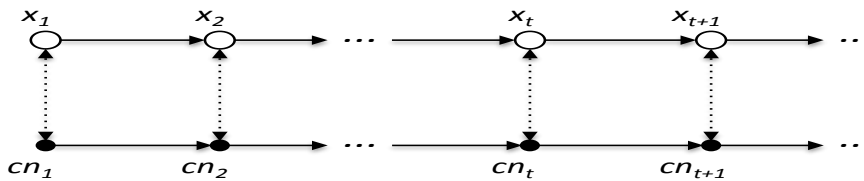

Figure 6: Transforming a high dimensional time series model ($x_t \in R^n$) to a univariate time series model, using their corresponding contextual numbers ($cn_t \in R$) as a continuous numerical field.

Having the converted univariate time series model, now one is able to select any approach for this univariate time series model. Therefore, for any high dimensional time series model, we perform the following steps. Since the main underlying idea of our proposed algorithm is based on the Contextual Numbers, from now on we simply call our proposed approach as CN algorithm.

1- Using $N$ high dimensional training data points, train a one dimensional SOM with $K$ nodes and transform those data points to $N$ corresponding contextual numbers.
2- Assuming a $d$-step time lag (equation 5) fit a model to the time series of $N$ contextual numbers.
3- Using the trained SOM, first project the new high dimensional observations to extract their most likely contextual numbers (as described in Section 3.1) and then predict the next step contextual numbers using the trained time series model in step 2.
4- Transform back the predicted contextual numbers to the high dimensional observation space, using the trained SOM in step 1, by selecting the appropriate high dimensional weight vector assigned to the predicted contextual number. Note that if the predicted contextual number is not an integer, one can simply estimate its weight vector by interpolating the weight vectors of its left and right contextual numbers (That is. indices of the trained SOM in step 1).

Although there are a wide variety of methods for the case of univariate time series forecasting (step 2), in this work we just applied a linear regression model to the time series of contextual numbers as it needs less parameters to optimize, but in general one can use any forecasting method for this step.

Using a linear regression model, there are only two free parameters (hyper-parameters) to optimize in CN algorithm. One is the number of nodes in the first one dimensional SOM, $K$, and the second parameter is $d$, the time lag in time series model (equation 5). Therefore, by performing cross validation and appropriate grid search one is able to find a relatively good set of values for the hyper-parameters.

In the next section we present the results of several experiments.

## 5. Experiments

In order to assess the quality of the proposed approach for high dimensional time series forecasting, we conducted several experiments. We compared the accuracy of our predictions to the prediction of LDS approach as implemented in (Ghahramani & Hinton, 1996). For each time slice, the relative Euclidean distance between estimated and the real high dimensional vectors, $e_t = \|x_t^{est} - x_t^{real}\| / \| x_t^{real}\|$ is calculated as the prediction error at time slice $t$.

In the first set of experiments, we used relatively low dimensional data sets, in order to compare our results with the result of LDS method.

We used 4 data sets from (Rogers *et al.* 2013) with the following descriptions:

1- SST: A sequence of 2000 Sea Surface Temperature (SST) observations, each as a 30 dimensional vectors (a grid of 5 by 6). Further, first1800 points were selected as training set and 200 points as test data.
2- Video: 1171 gray scale frames of ocean surface during low tide spatially down-sampled to a 10-by-10 patch for each frame, and normalized. Further, first 1000 points were selected as training set and 171 points as test data.
3- Tesla: Opening, closing, high, low, and volume of the stock prices of 12 car and oil companies, constructing a 60 dimensional vector at each time step, totally 724 time steps. First 700 time steps were selected as training data and the rest for testing.
4- NASDAQ-100: Opening, closing, adjusted-closing, high, low, and volume for 20 randomly chosen NASDAQ-100 companies, constructing a 120 dimensional vector for 1259 time steps. First 1200 were selected as training data and the rest for testing.

For both financial time series (Tesla and NASDAQ-100), we normalized each individual time series separately (For example. closing values of the first company) since unlike two other data sets, the values in different dimensions are not in the same range. For LDS, the training parameters are set as reported in (Rogers *et al.* 2013). In all of the cases, we made sure that LDS converges based on the log-likelihood criteria. For CN algorithm, we trained each one dimensional SOM based on batch training algorithm, which can be parallelized if necessary (Vesanto *et al.* 1999). However, we did not use parallel processing in none of the experiments. The training parameters of SOM algorithm were set as proposed in (Vesanto *et al.* 1999). For each case, we tested CN, only with 25 random choices for two hyper-parameters, where $K$ was selected randomly from the range of 30 to 200 and $d$ from the range of 2 to 20.

Figure 7 shows the results of our proposed method, in comparison with LDS method. As it is shown, two algorithms have the same pattern, while CN is performing better than LDS in all the cases, except video data set, where LDS outperforms CN partially.

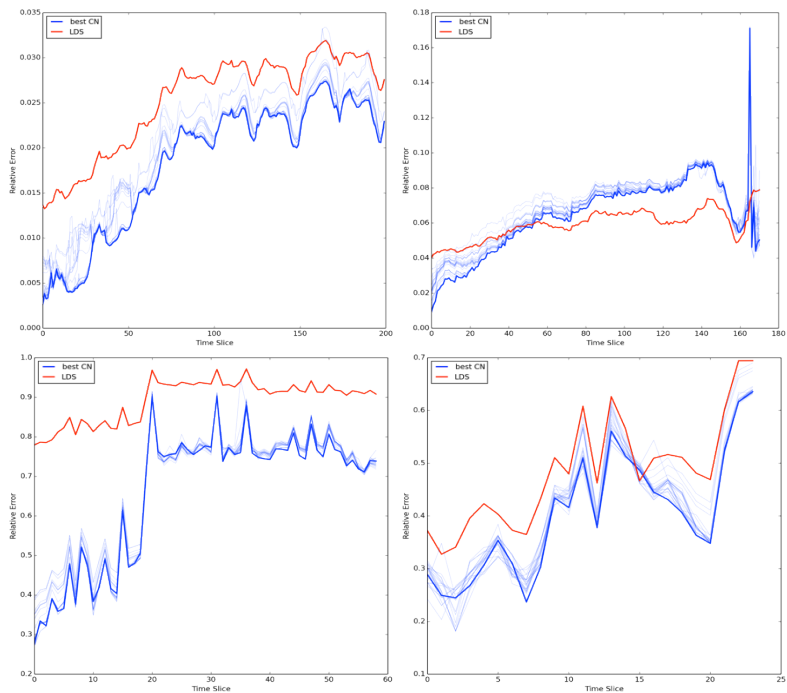

Figure 7: Forecasting results for 25 CN models with randomly selected hyper-parameters and LDS applied to SST data (top left) and video data (top right), NASDAQ-100 (bottom left) and Tesla (bottom right)

For the video data set CN starts with a better results, but gradually the quality of predictions get worse comparatively. As mentioned in Section 3.1 about projection of high dimensional data sets to contextual numbers, plotting the maximum posterior probabilities, $p_{max} = \max_i\bigl(p(cn_i|x)\bigr)$, for the video data set shows that there is an inverse relation between $p_{max}$ and the relative error of prediction. Figure 8 shows the pattern of posterior probabilities, calculated based on equation 2 for the testing data of video time series. As it is shown in Figure 8, when $p_{max}$ gradually decreases we have higher prediction errors (Figure 7 top right panel) and then as in the second half of the testing data (from time slice 140) $p_{max}$ increases slightly, it brings the forecasting errors lower. A decline in the posterior probability is indicating that the new observations do not find any distinctive indices in the trained one-dimensional SOM. Therefore in practice, by monitoring the posterior probabilities during the projection phase, one can define a threshold in the relative value of $p_{max}$ to trigger a retraining of SOM with new set of observations. Nevertheless, if the training data is large enough, training only one SOM should be sufficient, as large data set should include all the potential states in the high dimensional space.

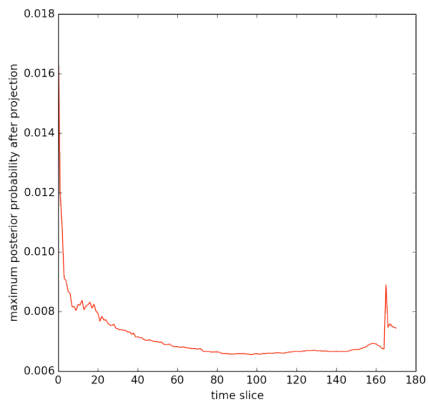

Figure 8: Maximum posterior probability of contextual numbers, $p_{max}$, for the test data in video data set

Although LDS is a powerful method, it becomes computationally expensive for high dimensional vectors (Horenko *et al.* 2008). On the other side in our proposed algorithm, since high dimensional states are encapsulated in a one dimensional numerical field, majority of calculations are done in a simple one dimensional data set. The only computational step, which happens only once, is to train a one dimensional SOM based on the high dimensional training data. Based on the conducted experiments in (Vesanto *et al.* 1999), SOM algorithm scales linearly with the dimensionality and the size of training data. Further, for speeding up the process it is possible to parallelize the training process.

In the second set of experiments, we compared the time efficiency of two approaches. We used NOAA high-resolution data of daily Sea Surface Temperature (SST) of Nino-3.0 region from 1998 to 2002, totally for 5 years (Reynolds *et al.* 2007). In the original data set, each data point has 8000 spatial dimensions in a grid of 40 by 200 nodes. In order to analyze the behavior of each method regarding the dimensionality of observation states, we spatially down-sampled the original observations to the following dimensions: 20, 80,125, 320, 500, 2000 and 8000.

We had the same set-ups for both LDS and CN methods as we did for the previous example. The Matlab code for LDS algorithm (Ghahramani & Hinton, 1996) did not converge for dimensionalities of 2000 and 8000. For 2000 dimensions after 1,384 seconds the log-likelihood criteria did not improve at all, so it was stopped automatically. On the other hand, as we expected, CN method performed relatively linearly in different resolutions. Figure 9 shows the time efficiency of both algorithms in comparison to each other.

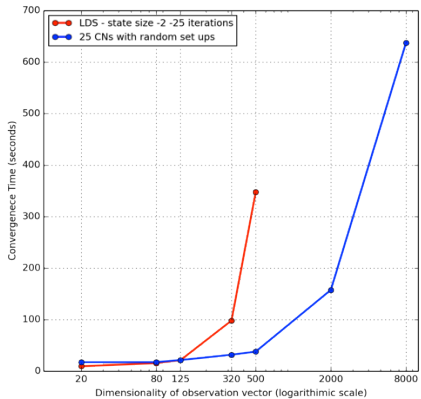

Figure 9: Training times of LDS and CN methods for different dimensionalities of observations

## 6. Conclusions

In this work, from a certain perspective to SOM algorithm, based on its less discussed functionality as an encoder and dimensionality reduction method, we proposed a methodology for transforming high dimensional data to a continuous one dimensional numerical field, where for each data point in the high dimensional space, there will be a corresponding scalar values, while similar values refer to similar points in the high dimensional space. We called this one-dimensional code, contextual number. It is a number since these codes can be used similar to usual numbers in any numerical calculations and further it is contextual, since unlike usual numerical values (For example. temperature measurements) with a-priori given interpretation (That is. a sense of temperature as being high or low), contextual numbers can be interpreted only based on their high dimensional contexts.

The automatic nonparametric mapping (done by a one dimensional SOM) from a high-dimensional space to one dimensional contextual numbers can be explained as a feature learning step, where the final representation (That is. contextual numbers) can be embedded in any data driven modeling task. In this work, we showed how contextual numbers could be used in the context of high dimensional spatiotemporal dynamics.

One potential direction for future applications of contextual numbers would be their application to the problem of multimodal learning such as image annotation problem, in which the ultimate goal is to find the joint probability distribution over a set of images and a bunch of texts (key words) as candidate labels for images. In this case, one idea would be that by replacing observations in each mode with their corresponding contextual numbers (where similar contextual numbers refer to similar images or similar words), the problem of multimodal learning would be converted to a classical function approximation over the two dimensional state space of contextual numbers.

Further, although we showed that contextual numbers could be used to convert a high-dimensional time series to a univariate time series model, there are many cases, in which the original time series consists of several modes of high dimensional time series with different contexts such as time series of temperature, wind speed and air pressure over a specific region, while one is interested in the integration of these different contexts. In this situations as well, contextual numbers (in two levels) can be used to transform the original time series to a time series of several distinct contextual numbers and then to another univariate time series model with another transformation.

## Acknowledgments
This work was established at the Singapore-ETH Centre for Global Environmental Sustainability (SEC), Future Cities Laboratory, cofounded by the Singapore National Research Foundation (NRF) and ETH Zurich.
## References
Barreto, G. A., and Araujo, A. F. Identification and control of dynamical systems using the self-organizing map. *Neural Networks, IEEE Transactions on*, *15*(5), 1244-1259. 2004.

Barreto, G. A., and Souza, L. G. M. Adaptive filtering with the self-organizing map: a performance comparison. *Neural Networks*, *19*(6), 785-798. 2006.

Bengio, Y., Courville, A., and Vincent, P. Representation learning: A review and new perspectives. *Pattern Analysis and Machine Intelligence, IEEE Transactions on*, *35*(8), 1798-1828. 2013.

Bishop, C. M., Svensén, M., and Williams, C. K. GTM: The generative topographic mapping. *Neural computation*, *10*(1), 215-234. 1998.

Cheng, Y. Convergence and ordering of Kohonen's batch map. *Neural Computation*, *9*(8), 1667-1676. 1997.

Erwin, E., Obermayer, K., and Schulten, K. Self-organizing maps: ordering, convergence properties and energy functions. *Biological cybernetics*, *67*(1), 47-55. 1992.

Flanagan, J. A. Self-organization in the one-dimensional SOM with a decreasing neighborhood. *Neural Networks*, *14*(10), 1405-1417. 2001.

Ghahramani, Z., and Hinton, G. E. *Parameter estimation for linear dynamical systems* (pp. 1-6). Technical Report CRG-TR-96-2, University of Totronto, Dept. of Computer Science, 1996.

Goerg, G. M., and Shalizi, C. R. LICORS: Light cone reconstruction of states for non-parametric forecasting of spatio-temporal systems. *arXiv preprint arXiv:1206.2398*. 2012.

Hinton, G. E., and Salakhutdinov, R. R. Reducing the dimensionality of data with neural networks. *Science*, *313*(5786), 504-507. 2006.

Hinton, G., Osindero, S., and Teh, Y. W. A fast learning algorithm for deep belief nets. *Neural computation*, *18*(7), 1527-1554. 2006.

Horenko, I., Dolaptchiev, S. I., Eliseev, A. V., Mokhov, I. I., and Klein, R. Metastable decomposition of high-dimensional meteorological data with gaps. *Journal of the Atmospheric Sciences*, *65*(11), 3479-3496. 2008.

Kiviluoto, K. Topology preservation in self-organizing maps. In *IEEE International Conference on Neural Networks* (Vol. 1, No. 1996, pp. 294-299). 1996.

Kohonen, T. Self-organized formation of topologically correct feature maps. *Biological cybernetics*, *43*(1), 59-69. 1982.

Kohonen, T. Improved versions of learning vector quantization. In *Neural Networks, 1990 IJCNN International Joint Conference on* (pp. 545-550). IEEE. 1990.

Kohonen, T., and Maps, S. O. Springer series in information sciences. *Self-organizing maps*, *30*. 1995.

Kohonen, T. Essentials of the self-organizing map. *Neural Networks*, *37*, 52-65. 2013.

Kostiainen, T., and Lampinen, J. On the generative probability density model in the self-organizing map. *Neurocomputing*, *48*(1), 217-228. 2002.

Kruskal, J. B. Multidimensional scaling by optimizing goodness of fit to a nonmetric hypothesis. *Psychometrika*, *29*(1), 1-27. 1964.

Luttrell, S. P. Derivation of a class of training algorithms. *Neural Networks, IEEE Transactions on*, *1*(2), 229-232. 1990.

Luttrell, S. P. A Bayesian analysis of self-organizing maps. *Neural Computation*, *6*(5), 767-794. 1994.
14